\documentclass[11pt]{article}
\usepackage{soul}
\usepackage{epsfig,subfigure}
\usepackage{amssymb, amsmath}
\usepackage{color} 
\usepackage{hyperref}  
\usepackage{xcolor,cancel}
\usepackage{graphicx}
\usepackage{color}

\usepackage{amssymb}
\usepackage{amsmath,amsfonts,amssymb}

\usepackage{verbatim}
\usepackage{stfloats}
\usepackage[bookmarks=false]{}

\newtheorem{theorem}{Theorem}

\newtheorem{definition}{Definition}
\newtheorem{lemma}{Lemma}

\newtheorem{remark}{Remark}

\begin{document}
\date{}
\title{$K$-User Symmetric $M\times N$ MIMO Interference Channel under Finite Precision CSIT: A GDoF perspective}
\author{ \normalsize Arash Gholami Davoodi and Syed A. Jafar \\
{\small Center for Pervasive Communications and Computing (CPCC)}\\
{\small University of California Irvine, Irvine, CA 92697}\\
{\small \it Email: \{gholamid, syed\}@uci.edu}
}
\maketitle

\begin{abstract}
Generalized Degrees of Freedom (GDoF) are characterized for the symmetric $K$-user Multiple Input Multiple Output (MIMO) Interference Channel (IC) under the assumption that the channel state information at the transmitters (CSIT) is limited to finite precision. In this symmetric setting, each transmitter is equipped with $M$ antennas, each receiver is equipped with $N$ antennas, each desired channel (i.e., a channel between a transmit antenna and a receive antenna belonging to the same user) has strength $\sim P$, while each undesired channel has strength $\sim P^\alpha$, where $P$ is a nominal SNR parameter.  The result generalizes a previous GDoF characterization for the SISO setting $(M=N=1)$ and is enabled by a significant extension of the Aligned Image Sets bound that is broadly useful. GDoF per user take the form of a $W$-curve with respect to $\alpha$ for fixed values of $M$ and $N$. Under finite precision CSIT, in spite of the presence of multiple antennas, all the benefits of  interference alignment are lost.
\end{abstract}

\section{Introduction}
Much of the progress in our understanding of the capacity limits of wireless networks over the past decade has come from the pursuit of progressively refined capacity approximations. Generalized degrees of freedom  (GDoF) characterizations represent a most significant step along this path because of their ability to capture arbitrary channel strength and channel uncertainty levels.  The GDoF framework may seem counter-intuitive at first because it allows exponential scaling of signal strengths with various exponents. An intuitive justification for the GDoF framework is as follows. It is important to remember that the goal behind GDoF is to seek capacity approximations for a given wireless network with its arbitrary \emph{finite} signal strengths and channel uncertainty levels. Unlike the degrees of freedom (DoF) metric which linearly scales all signal strengths and loses the distinction of different channel strengths (every non-zero channel carries $1$ DoF), the GDoF formulation takes a more sophisticated approach.  The key to GDoF is the intuition that if the capacity of every link in a network is scaled by the same factor, then the capacity region of the network should scale by approximately the same factor as well. Normalizing the capacity of the network by the scaling factor then yields a capacity approximation for the original network. Following this intuition, one allows the scaling factor to approach infinity, while guaranteeing that the capacity is always normalized by the scaling factor. The asymptotic behavior of normalized capacity is potentially easier to characterize than a direct approximation of the capacity of the original network. Let $\alpha_i$ represent the capacity of the $i^{th}$ link in the original network (in isolation from all other links), and let $\log(P)$ be the scaling factor applied to every link capacity. Then we obtain channels whose capacity scales as $\alpha_i\log(P)$, i.e., channels whose strength scales as $P^{\alpha_i}$, and, according to this intuitive reasoning, normalization of network capacity by $\log(P)$ in the limit $P\rightarrow\infty$ presents the approximation of the capacity of the original network. This approximation is what is known as the GDoF characterization, and along with its abstractions into deterministic channel models, over the past decade it has been the key to finding capacity approximations for many networks whose exact capacity remains intractable. Thus, the linear scaling of capacity naturally corresponds to an exponential scaling of signal strengths in the GDoF model.

\begin{figure*}[h]
\center
\begin{minipage}[c]{0.54\textwidth}
	\centerline{\includegraphics[width=3.7in]{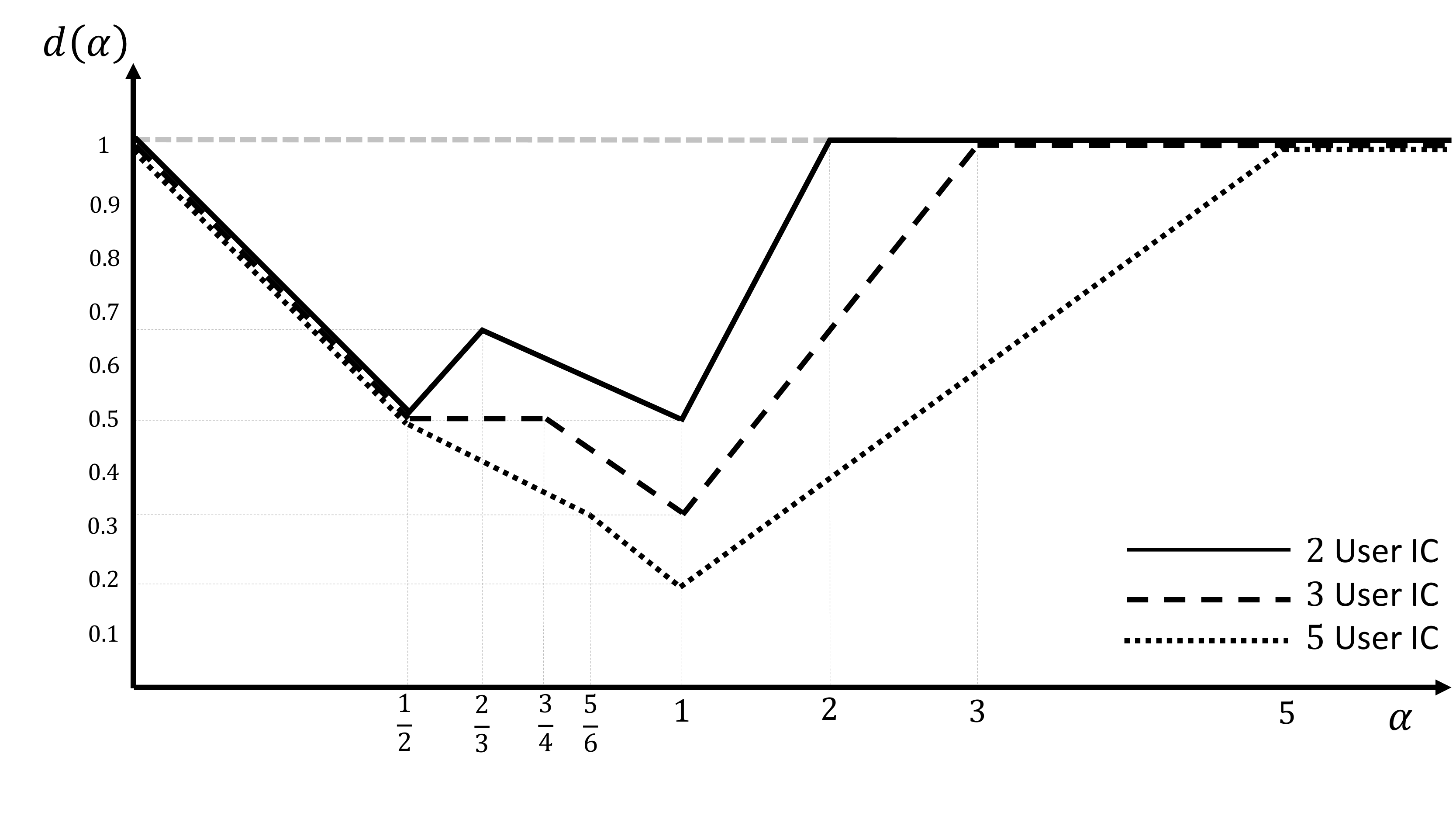}}
\end{minipage}~~~~~
\begin{minipage}[c]{0.33\textwidth}
\begin{eqnarray*}
d(\alpha) = \left\{
\begin{array}{ll} 
1-\alpha,&0\leq\alpha\leq \frac{1}{2}\vspace{0.03in}\\ 
\frac{K-2-(K-3)\alpha}{K-1}, &\frac{1}{2}<\alpha\leq\frac{K}{K+1}\vspace{0.03in}\\ 
1-\left(\frac{K-1}{K}\right)\alpha,&\frac{K}{K+1}<\alpha\leq 1\\
\frac{\alpha}{K},& 1<\alpha\leq K\\
1, &K<\alpha
\end{array}
\right.
\end{eqnarray*}
\end{minipage}
\caption{GDoF/user of the Symmetric $K$ User Interference Channel with Finite Precision CSIT \cite{Arash_Jafar_IC}.}\label{fig:Kusert}
\end{figure*}

GDoF studies started with settings where perfect CSIT is available \cite{Etkin_Tse_Wang, Jafar_Vishwanath_GDOF, Karmakar_Varanasi}. The opposite extreme of no CSIT was also explored under strong assumptions of statistical equivalence between users \cite{Huang_Jafar_Shamai_Vishwanath, Varanasi_noCSIT, Guo_noCSIT}. Lately, however, the focus has shifted to the broader assumption of finite precision CSIT  \cite{Arash_Jafar_TC}, \cite{Arash_Bofeng_Jafar_BC}. Some of the more sophisticated concepts such as interference alignment \cite{Jafar_FnT} have turned out to be too fragile to be useful with finite precision CSIT, so that conventional achievable schemes are usually optimal. As such the  main challenge for GDoF studies under finite precision CSIT tends to be the proof of optimality, i.e., the converse, or the GDoF outer bound.  Finding tight GDoF outer bounds under finite precision CSIT is generally a hard problem, as exemplified by the conjecture of Lapidoth et al. \cite{Lapidoth_Shamai_Wigger_BC} which remained unresolved for nearly a decade. The main idea for these outer bounds is the Aligned Image Sets (AIS) argument that was introduced in  \cite{Arash_Jafar_PN} in order to settle the conjecture of Lapidoth et al. Generalizations of the AIS approach have also helped settle the GDoF in other settings such as the X channel and the $2$ user MISO BC under finite precision CSIT in \cite{Arash_Jafar_TC}, and the 2 user MISO BC with arbitrary channel strengths and channel uncertainty levels in \cite{Arash_Bofeng_Jafar_BC}. Of particular relevance to this work is  \cite{Arash_Jafar_IC} where the sum GDoF of $K$-user symmetric interference channel (IC) is characterized under finite precision CSIT (see Figure \ref{fig:Kusert}). This work is motivated by the goal of further broadening the scope of the AIS argument, so that the results of \cite{Arash_Jafar_IC} may be generalized to MIMO settings.

In this paper, we characterize the GDoF for the symmetric  $K$-user MIMO Interference Channel  under the assumption that the CSIT is limited to finite precision. In this symmetric setting, each transmitter is equipped with $M$ antennas, each receiver is equipped with $N$ antennas, each desired channel (i.e., a channel between a transmit antenna and a receive antenna belonging to the same user) has strength $\sim P$, while each undesired channel has strength $\sim P^\alpha$, where $P$ is a nominal SNR parameter.  GDoF per user take the form of a $W$-curve with respect to $\alpha$ for fixed values of $M$ and $N$. See Figure \ref{Fig1}. As usual for finite precision CSIT, achievability is fairly straightforward. While ostensibly the main result of this work is the GDoF characterization for the $K$-user symmetric MIMO IC, the deeper significance of this paper resides in a key generalization of the AIS approach that allows comparisons in the GDoF sense of the entropies of  different numbers of linear combinations (finite precision versus perfectly known channels) of random variables under various power-level partitions. The generalization seems broadly useful for GDoF problems related to MIMO wireless networks.

{\it Notation:} The notation $|A|$ denotes the cardinality of the set $A$ and the notation $[n]$ is defined as $\{1,2,\cdots,n\}$ for any $n\in\mathbb{N}$ where $\mathbb{N}$ is the set of all positive integer numbers. The notations~ $X^{[T]}$ and $X_{i}^{[T]}$ also stand~ for $\{X(1), X(2), \cdots X(T)\}$ and $\{X_i(t): \forall t\in[T]\}$, respectively. Moreover, we use the Landau $o(\cdot)$ notation for the functions $f(x), g(x)$ from $\mathbb{R}$ to $\mathbb{R}$ as follows. $f(x)=o(g(x))$ denotes that $\limsup_{x\rightarrow\infty}\frac{|f(x)|}{|g(x)|}=0$. Finally, we define $\lfloor x\rfloor$  as the largest integer that is smaller than or equal to $x$  for any positive real number $x$ and  the smallest integer that is larger than or equal to $x$  for any negative real number $x$. $A^\dagger$ is the transpose of matrix $A$. The support of a random variable $X$ is denoted as supp$(X)$.


\section{Definitions}
\begin{definition}\label{bd}[Bounded Density Channel Coefficients \cite{Arash_Jafar_PN}] Define a set of real valued random variables, $\mathcal{G}$ such that the magnitude of each random variable $g\in\mathcal{G}$ is bounded away from  infinity, $ |g|\leq\Delta<\infty$, for some positive constant $\Delta$, and there exists a finite positive constant $f_{\max}$, such that for all finite cardinality disjoint subsets $\mathcal{G}_1, \mathcal{G}_2$ of $\mathcal{G}$, the joint probability density function of all random variables in $\mathcal{G}_1$, conditioned on all random variables in $\mathcal{G}_2$, exists and is bounded above by $f_{\max}^{|\mathcal{G}_1|}$. Without loss of generality we will assume that $f_{\max}\geq 1, \Delta\geq 1$.
\end{definition}

\begin{definition}[Power Levels] Consider integer valued random variables $X_i$ over alphabet $\mathcal{X}_{\lambda_i}$,
\begin{eqnarray}
\mathcal{X}_{\lambda_i}&\triangleq&\{0,1,2,\cdots,\bar{P}^{\lambda_i}-1\}
\end{eqnarray}
where $\bar{P}^{\lambda_i}$ is a compact notation for $\left\lfloor\sqrt{P^{\lambda_i}}\right\rfloor$ and the constant $\lambda_i$ is a positive real number denoting the \emph{power level} of $X_i$. 
\end{definition}

\begin {definition}\label{powerlevel} For $X\in\mathcal{X}_\lambda$, and $0\leq \lambda_1\leq\lambda$, define the random variables $(X)_{\lambda_1}^{\lambda}$ as,
 \begin{eqnarray}
(X)_{\lambda_1}^{\lambda}&\triangleq&\left \lfloor \frac{X}{\bar{P}^{\lambda_1}} \right \rfloor
\end{eqnarray}
\end {definition}
In words, $(X)_{\lambda_1}^{\lambda}$ retrieves the top $\lambda-\lambda_1$ power levels of $X$. Similarly, for  the vector ${\bf V}=\begin{bmatrix}v_1&v_2&\cdots&v_k\end{bmatrix}^\dagger$, we define $({\bf V})^{\lambda}_{\lambda_1}$ as,
 \begin{eqnarray}
({\bf V})^{\lambda}_{\lambda_1}&\triangleq&\begin{bmatrix}(v_1)^{\lambda}_{\lambda_1}&(v_2)^{\lambda}_{\lambda_1}&\cdots&(v_k)^{\lambda}_{\lambda_1}\end{bmatrix}^\dagger
\end{eqnarray}

\begin {definition}\label{deflc} For {real}  numbers $x_1,x_2,\cdots,x_k{\in\mathcal{X}_{\eta}}$ define the notation $L_j^b(x_i,1\le i\le k)$  to represent,
\begin {eqnarray}
L^b_j(x_1,x_2,\cdots,x_k )&=&\sum_{1\le i\le k} \lfloor g_{j_i}x_i\rfloor
\end{eqnarray}
for  distinct random variables $g_{j_i}\in\mathcal{G}$. The subscript $j$ is used to distinguish among multiple linear combinations, and may be dropped if there is no potential for ambiguity.  For  the vector $V=\begin{bmatrix}v_1&v_2&\cdots&v_k\end{bmatrix}^\dagger$ define the notation $L^b_j(V)$  to represent,
\begin {eqnarray}
L^b_j(V)=\sum_{1\le i\le k} \lfloor g_{j_i}v_i\rfloor
\end{eqnarray}
for  distinct random variables $g_{j_i}\in\mathcal{G}$.
\end {definition}

\begin{definition}\label{defvec} For the two vectors $V=\begin{bmatrix}v_1&\cdots&v_{k_1}\end{bmatrix}^\dagger$ and $W=\begin{bmatrix}w_1&\cdots&w_{k_2}\end{bmatrix}^\dagger$ define the vector $V\bigtriangledown W$ as $\begin{bmatrix}v_1&\cdots&v_{k_1}&w_1&\cdots&w_{k_2}\end{bmatrix}^\dagger$.
\end{definition}

\section{System Model} {\label{sec-sys}}
In this work we consider only the setting where all variables take real values. Extensions to complex settings are cumbersome but conceptually straightforward  as in \cite{Arash_Jafar_PN}.
\subsection{The Channel}
\begin{figure}[tp]
\centering 
\includegraphics[scale =0.28]{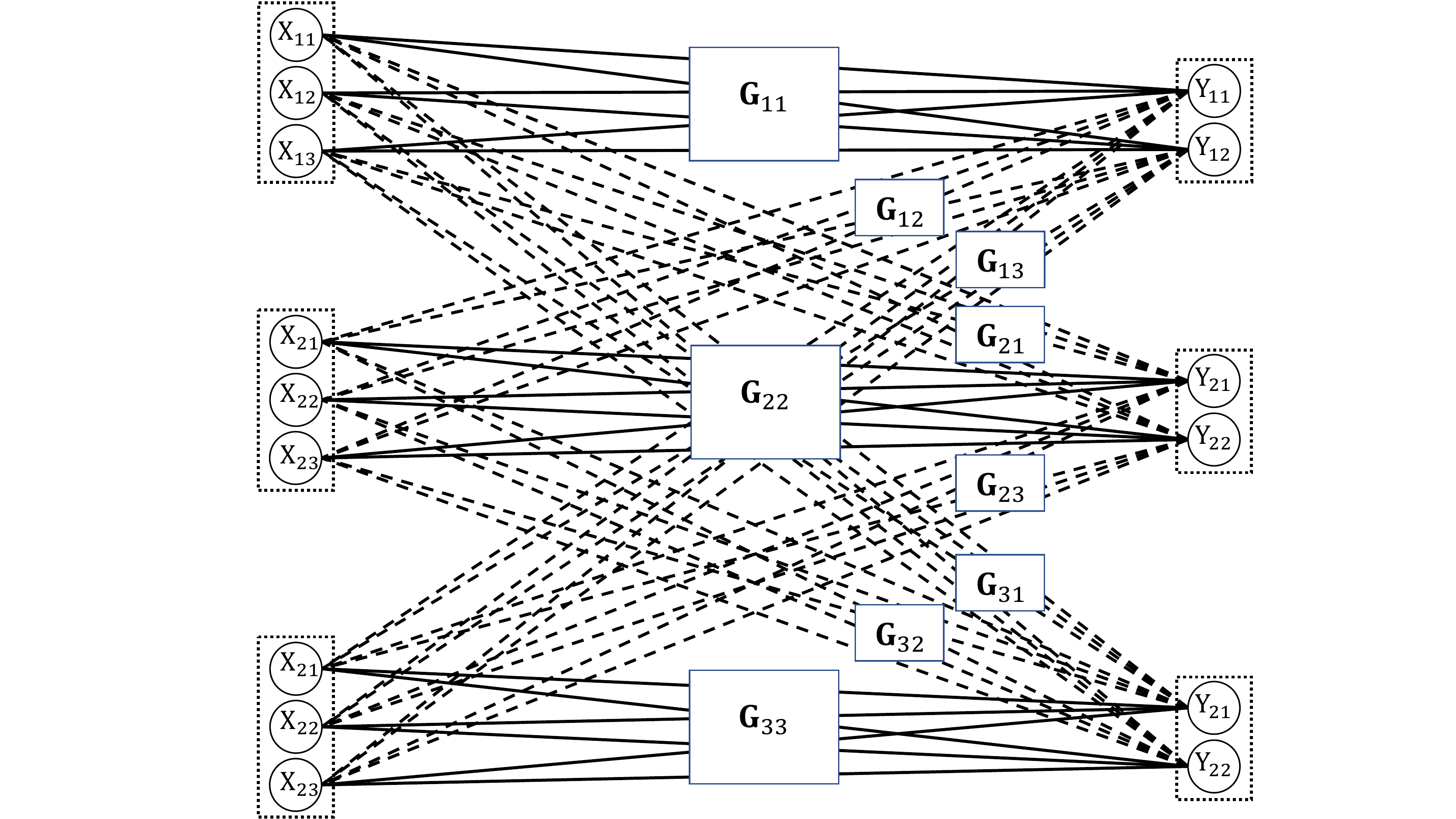}
\caption{Three user $3 \times 2$ MIMO IC.}
\label{Fig1rr}
\end{figure}

\begin{figure}[tp]
\centering 
\includegraphics[scale =0.4]{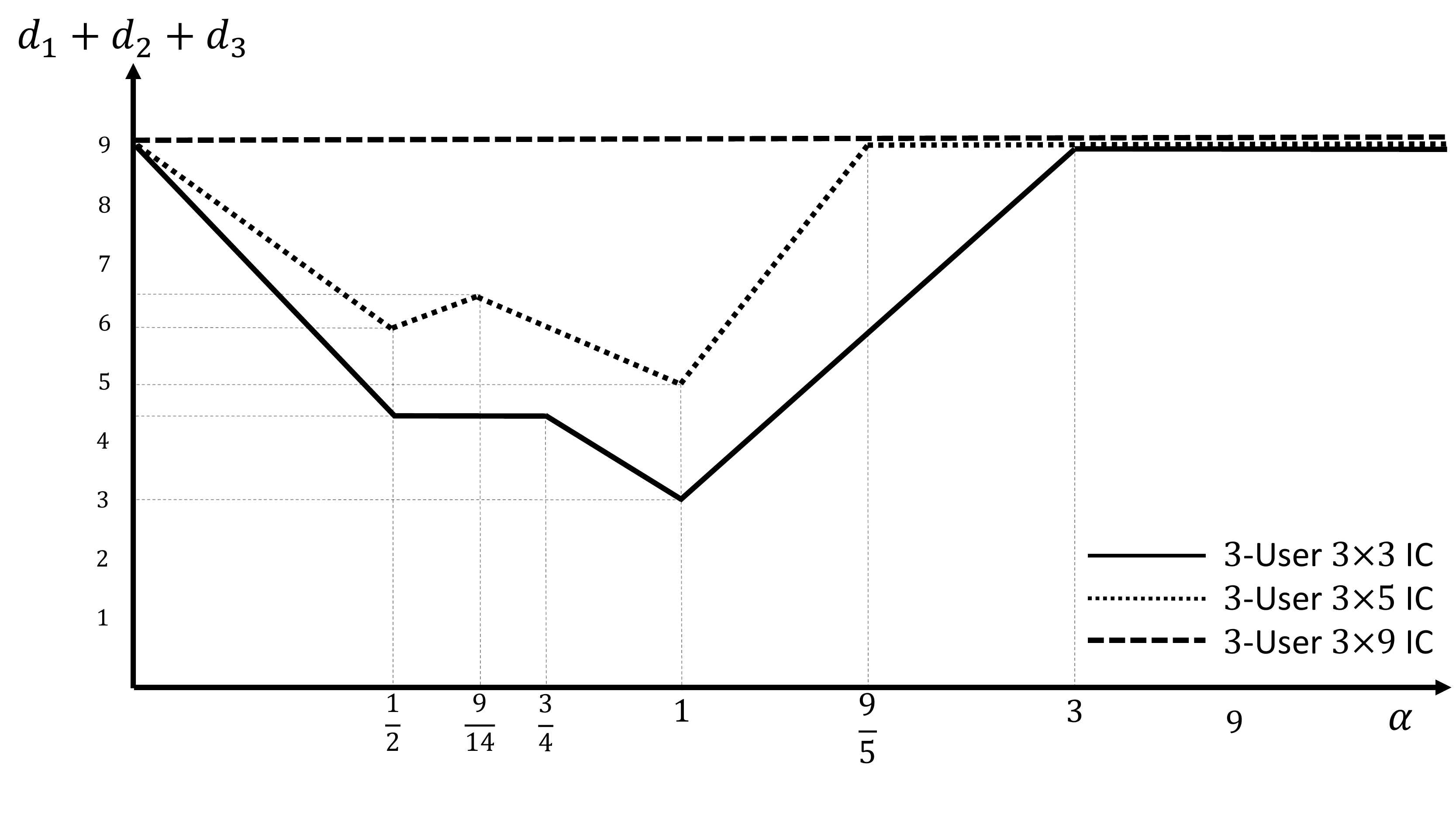}
\caption{Sum GDoF of the three user $3 \times N$ MIMO IC.}
\label{Fig1}
\end{figure}

\noindent Define  random variables $\mathbf{X}_{k}(t)$ and $\mathbf{Y}_{k}(t)$, $\forall k\in[K]$ as,
\begin{align}
\mathbf{X}_{k}(t)=&\begin{bmatrix}{X}_{k1}(t)&{X}_{k2}(t)&\cdots&{X}_{kM}(t)\end{bmatrix}^\dagger\\
\mathbf{Y}_{k}(t)=&\begin{bmatrix}{Y}_{k1}(t)&{Y}_{k2}(t)&\cdots&{Y}_{kN}(t)\end{bmatrix}^\dagger
\end{align}
where the channel uses are indexed by $t\in[T]$. $X_{km}(t), k\in[K],m\in[M],t\in[T]$ are the symbols sent from $m$-th transmit antenna of the $k$-th transmitter and are subject to unit power constraint, while $Y_{kn}(t), k\in[K],n\in[N],t\in[T]$ are the symbols observed by the $n$-th antenna of the $k$-th receiver. Under the GDoF framework, the channel model for the $K$-user MIMO IC is defined by the following input-output equations
\begin{align}
{\bf Y}_{k}(t)=&\sqrt{P}{\bf G}_{kk}(t)\mathbf{X}_k(t)+\sqrt{{P}^{\alpha}}\sum_{\hat{k}=1,\hat{k}\neq k}^K{\bf G}_{k\hat{k}}(t)\mathbf{X}_{\hat{k}}(t)+{\bf \Gamma}_{k}(t)
\end{align}
for all $k\in[K]$ and $t\in[T]$. The $N\times M$ matrix ${\bf G}_{k\hat{k}}(t)$ is the channel fading coefficient matrix between the $k$-th receiver and the $\hat{k}$-th transmitter for any $k,\hat{k}\in[K]$. The entry in the $n$-th row and $m$-th column of the matrix ${\bf G}_{k\hat{k}}(t)$ is ${G}_{k\hat{k}nm}(t)$. $\mathbf{\Gamma}_{k}(t)$ are  $N\times 1$  matrices whose components are zero mean unit variance additive white Gaussian noise (AWGN) experienced by $k$-th receiver.  Figure \ref{Fig1rr} illustrates a $3$-user $3\times 2$ MIMO IC.  $P$ is a nominal SNR parameter that approaches infinity for GDoF characterizations. CSIR is assumed to be perfect. However, CSIT is limited to finite precision. Under finite precision CSIT  we assume that $G_{k\hat{k}nm}(t)\in\mathcal{G}$ for any $k,\hat{k}\in[K],n\in[N],m\in[M]$ and $t\in[T]$, and since transmitters only know the probability density but not the realizations of channel coefficients, we assume that all ${\bf X}_k(t), t\in[T], k\in[K]$ are independent of $\mathcal{G}$.
\subsection{GDoF}
The definitions of achievable rates $R_i(P)$ and capacity region $\mathcal{C}(P)$ are standard. The GDoF region is defined as
\begin{eqnarray}
\mathcal{D}&=&\{(d_1,d_2,\cdots,d_K): \exists (R_1(P),R_2(P), \cdots,R_K(P))\nonumber\\
&&\in\mathcal{C}(P), \mbox{ s.t. } d_k=\lim_{P\rightarrow\infty}\frac{R_k(P)}{\frac{1}{2}\log(P)}, \forall k\in[K]\}
\end{eqnarray}
The maximum value of $d_1+d_2+\cdots+d_K$ over $\mathcal{D}$ is known as the sum GDoF value.

\section{Main Result}

\begin{theorem}\label{theorem:GDoF1} The sum GDoF value for the $K$-user  symmetric MIMO IC for $M\le \frac{N}{K}$ is $KM$, and for $\frac{N}{K}\le M$ is
\begin{align}
{\sum_{k=1}^Kd_k}=&\left\{
\begin{array}{ll} 
K\min(M,N)(1-\alpha)+\frac{K(N-M)^+\alpha}{K-1},&0\leq\alpha\leq \frac{1}{2}\vspace{0.03in}\\ 
\min\left(\frac{K}{K-1}\left((K-2)\min(M,N)(1-\alpha)+N(\alpha)\right)\right., &\vspace{0.03in}\\ 
N\alpha+K\min(M,N)\left(1-\alpha\right)\big),&\frac{1}{2}<\alpha\leq 1\\
\min\left(D(\alpha),K\min(M,N)\right), &1<\alpha
\end{array}
\right.\label {fg<}
\end{align}
where $N(\alpha)$ and $D(\alpha)$ are defined as,
\begin{eqnarray}
N(\alpha)&=&\min((K-1)M,N)\alpha+(N-(K-1)M)^+(1-\alpha)\\
D(\alpha)&=&(N-(K-1)M)^++\min(N,(K-1)M)\alpha 
\end{eqnarray}
\end{theorem}
\begin{remark} The sum GDoF, i.e., \eqref{fg<} for $N< M$ yields,
\begin{align}
{\sum_{k=1}^Kd_k}=&{KN}\times  \left\{
\begin{array}{ll} 
(1-\alpha),&0\leq\alpha\leq \frac{1}{2}\vspace{0.03in}\\ 
\frac{K-2-(K-3)\alpha}{K-1}, &\frac{1}{2}<\alpha\leq\frac{K}{K+1}\vspace{0.03in}\\ 
1-(\frac{K-1}{K})\alpha,&\frac{K}{K+1}<\alpha\leq 1\\
\frac{\alpha}{K},& 1<\alpha\leq K\\
1, &K<\alpha
\end{array}
\right.\label {GDOFM>N}
\end{align}
\end{remark}

\section{Proof of Theorem \ref{theorem:GDoF1}: Converse}
 The first step in the converse proof, identical to \cite{Arash_Jafar_IC},
 is the transformation
into a deterministic setting such that a GDoF outer bound
on the deterministic setting is also a GDoF outer bound on
the original setting.  We start directly from the deterministic
model.

\subsection{Deterministic Model}\label{DM_1}
\vspace{-1em}
\begin{align}
\bar{\mathbf{Y}}_{k}(t)&=[\bar{Y}_{k1}(t)\ \bar{Y}_{k2}(t)\ \cdots\ \bar{Y}_{kN}(t)]^\dagger\\
\bar{Y}_{kn}(t)&=L_{kn1}^b(t)\left({(\bar{\mathbf{X}}_{k}(t))}^{\max(1,\alpha)}_{\max(1,\alpha)-1}\right)+L_{kn2}^b(t)\left({(\bar{\mathbf{X}}_{j}(t))}^{\max(1,\alpha)}_{\max(1,\alpha)-\alpha},\forall j\in[K],j\neq k\right)\label{dm1}
\end{align}
for all $k\in[K],n\in[N],t\in[T]$. $\bar{\mathbf{X}}_{k}(t)$ are  defined as,
\begin{align}
\bar{\mathbf{X}}_{k}(t)=&[\bar{X}_{k1}(t)\ \bar{X}_{k2}(t)\ \cdots\ \bar{X}_{kM}(t)]^\dagger\label{ggf1}
\end{align}
for any $k\in[K]$, $t\in[T]$ where $\bar{X}_{km}(t)\in\{0, 1, \cdots, {\bar{P}}^{\max(1,\alpha)}-1\}$, $\forall k\in[K],m\in[M],t\in[T]$. 

\subsection{ Key Lemma} 
The following  lemma is the critical generalization of the AIS bound needed for Theorem \ref{theorem:GDoF1}.
\begin{lemma}\label{lemma} Define the two random variables $\bar{\bf U}_1$ and  $\bar{\bf U}_2$ as,
\begin{eqnarray}
\bar{\bf U}_1&=&\left({U}_{11}^{[T]},{U}_{12}^{[T]},\cdots,{U}_{1N_1}^{[T]}\right)\label{lemmamimox1}\\
\bar{\bf U}_2&=&\left({U}_{21}^{[T]},{U}_{22}^{[T]},\cdots,{U}_{2N_2}^{[T]}\right)\label{lemmamimox2}
\end{eqnarray}
where for any $t\in[T]$, $U_{1n}(t)$ and $U_{2n}(t)$ are defined as,
\begin{eqnarray}
U_{1n}(t)&=&L_{1n}^b(t)\left((\bar{\mathbf{V}}_1(t))^{\eta}_{\eta-\lambda_{11}}\bigtriangledown(\bar{\mathbf{V}}_2(t))^{\eta}_{\eta-\lambda_{12}}\bigtriangledown\cdots\bigtriangledown(\bar{\mathbf{V}}_l(t))^{\eta}_{\eta-\lambda_{1l}}\right), \forall n\in[N_1]\label{lemmamimox3}\\
U_{2n}(t)&=&L_{2n}^b(t)\left((\bar{\mathbf{V}}_1(t))^{\eta}_{\eta-\lambda_{21}}\bigtriangledown(\bar{\mathbf{V}}_2(t))^{\eta}_{\eta-\lambda_{22}}\bigtriangledown\cdots\bigtriangledown(\bar{\mathbf{V}}_l(t))^{\eta}_{\eta-\lambda_{2l}}\right), \forall n\in[N_2]\label{lemmamimox4}
\end{eqnarray}
where $\bar{\mathbf{V}}_i(t)=\begin{bmatrix}\bar{V}_{i1}(t)&\cdots&\bar{V}_{iM_i}(t)\end{bmatrix}^\dagger$, $\bar{V}_{im}(t)\in\mathcal{X}_{\eta}$  are all independent of $\mathcal{G}$, and $0\le\lambda_{1i},\lambda_{2i}\le\eta$ for any $i\in[l]$.  Without loss of generality,  $(\lambda_{1i}-\lambda_{2i})^+$ are sorted in descending order, i.e., $(\lambda_{1i}-\lambda_{2i})^+\ge(\lambda_{1j}-\lambda_{2j})^+$ if $1\le i< j\le l$.  Then, for any acceptable\footnote{Let $\mathcal{G}(Z)\subset\mathcal{G}$ denote the set of all bounded density channel coefficients that appear in $\bar{\bf U}_1,\bar{\bf U}_2$.  $W$ is acceptable if  conditioned on any $\mathcal{G}_o\subset (\mathcal{G}/\mathcal{G}(Z))\cup \{W\}$, the channel coefficients $\mathcal{G}(Z)$ satisfy the bounded density assumption. For instance, any random variable $W$ independent of $\mathcal{G}$ can be utilized in  Lemma \ref{lemma}.} random variable ${W}$, if $N_1\le \min(N_2, \sum_{i=1}^lM_i)$ we have,
\begin{eqnarray}
&&H({\bar{\bf U}}_1\mid {W},\mathcal{G})-H({\bar{\bf U}}_2\mid {W},\mathcal{G})\nonumber\\
&\le&T\big((N_1-\sum_{i=1}^sM_i)(\lambda_{1,s+1}-\lambda_{2,s+1})^++\sum_{i=1}^sM_i(\lambda_{1i}-\lambda_{2i})^+\big)\log{\bar{P}}+T~o~(\log{\bar{P}})\label{lemmamimox5}
\end{eqnarray}
where $s$ must satisfy the condition  $\sum_{i=1}^{s}M_i\le N_1< \sum_{i=1}^{s+1}M_i$.  
\end{lemma} 
Proof of Lemma \ref{lemma} is based on the AIS argument and is relegated to Appendix \ref{lemmap}.

\subsection{Some Insights For the Three User $2\times3$ MIMO IC}
\begin{figure}[!h]
\centerline{\includegraphics[width=6in]{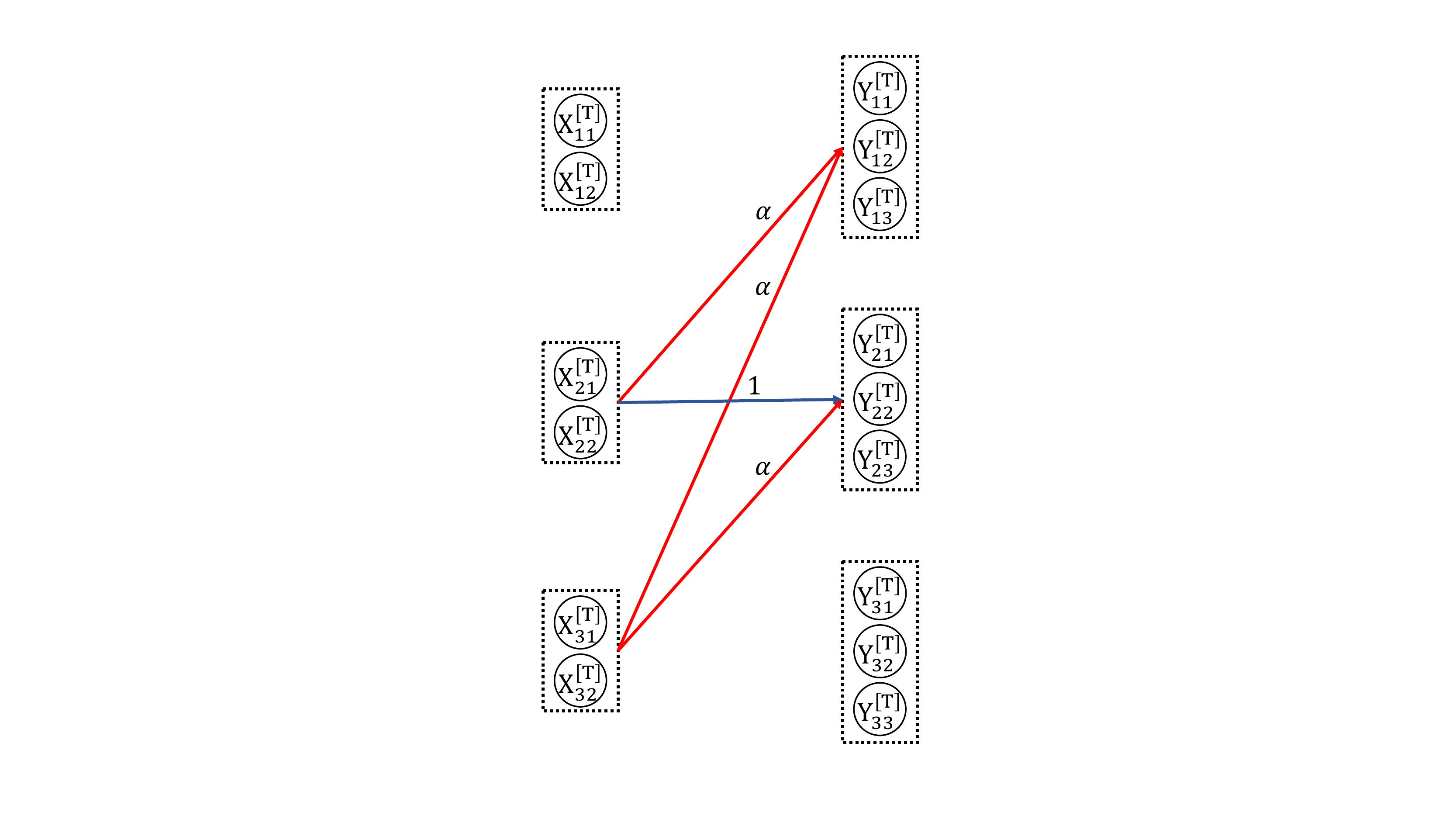}}
\caption{ Three user $2\times3$ MIMO IC. The network is fully connected but only the channel strength parameters needed for the application of Lemma \ref{lemma} are shown in this figure.}\label{fig:intuit-}
\end{figure}
To gain some insights into the application of Lemma \ref{lemma}, consider the three user $2\times3$ MIMO IC illustrated in Figure \ref{fig:intuit-} for $\alpha\le1$. To apply Lemma \ref{lemma}, the random variables  $\bar{\bf U}_1$, $\bar{\bf U}_2$, $\bar{\bf V}_1^{[T]}$, $\bar{\bf V}_2^{[T]}$ and $W$ are interpreted as $\bar{\bf Y}_2^{[T]}$, $\bar{\bf Y}_1^{[T]}$, $\bar{\bf X}_2^{[T]}$, $\bar{\bf X}_3^{[T]}$ and $\bar{\bf X}_1^{[T]}$, respectively. The first user receives the top $\alpha$ power levels of $\bar{\bf X}_2^{[T]}$ and $\bar{\bf X}_3^{[T]}$ while second reciever sees  the top $1$ power levels of $\bar{\bf X}_2^{[T]}$ and  the top $\alpha$ power levels of $\bar{\bf X}_3^{[T]}$.  So we have $\eta=1, \lambda_{11}=1, \lambda_{21}=\alpha, \lambda_{12}=\alpha, \lambda_{22}=\alpha$. Therefore, $(\lambda_{11}-\lambda_{21})^+=1-\alpha$ and $(\lambda_{12}-\lambda_{22})^+=0$. From Lemma \ref{lemma} we conclude,
\begin{align}
&H(\bar{\bf Y}_2^{[T]}\mid \bar{\bf X}_1^{[T]},\mathcal{G})-H(\bar{\bf Y}_1^{[T]}\mid \bar{\bf X}_1^{[T]},\mathcal{G})\nonumber\\
\le&T\big(1\times 0+2\times(1-\alpha)\big)\log{\bar{P}}+T~o~(\log{\bar{P}})\label{jl0}
\end{align}
  Let us also explain how intuitively we expect \eqref{jl0} to be true as well. Conditioned on $\bar{\bf X}_{1}^{[T]}$, $\bar{\bf Y}_{2}(t)$ is a linear combination of $\bar{\bf X}_{2}(t)$ and $(\bar{\bf X}_{3}(t))^{\alpha}$ while $\bar{\bf Y}_{1}(t)$ is a linear combination of $(\bar{\bf X}_{2}(t))^{\alpha}$ and $(\bar{\bf X}_{3}(t))^{\alpha}$. Consider the channel illustrated in Figure \ref{fig:intuit-}. First of all, observe that $\bar{\bf X}_{2}(t)$ appears in $\bar{\bf Y}_{2}(t)$ with the signal strength levels $1$ and appears in $\bar{\bf Y}_{1}(t)$ with the signal strength levels $\alpha$. Thus, due to the bounded density assumption the maximum difference of $2(1-\alpha)$ is possible in the GDoF sense between the two entropies. Note that, $\bar{\bf X}_{3}(t)$ appears in both the received signals  $\bar{\bf Y}_{1}(t)$ and  $\bar{\bf Y}_{2}(t)$ with the same signal strength levels of $\alpha$. Therefore, it cannot contribute positive difference of entropies as in the finite precision CSIT no interference alignment is possible.
  
Similarly, from Lemma \ref{lemma} we have,  
\begin{align}
H(\bar{\bf Y}_3^{[T]}\mid \bar{\bf X}_1^{[T]},\bar{\bf X}_2^{[T]},\mathcal{G})-H(\bar{\bf Y}_2^{[T]}\mid \bar{\bf X}_1^{[T]},\bar{\bf X}_2^{[T]},\mathcal{G})\le&2T(1-\alpha)\log{\bar{P}}+T~o~(\log{\bar{P}})\label{jl00}
\end{align} 
On the other hand, writing Fano's inequality for all the three users (and suppressing $o(T)$ terms for simplicity) we obtain the following bounds,
\begin{eqnarray}
TR_1&\le& H(\bar{\bf Y}_{1}^{[T]}\mid \mathcal{G})- H(\bar{\bf Y}_{1}^{[T]}\mid \bar{\bf X}_{1}^{[T]},\mathcal{G})\label{jl1}\\
TR_2&\le& H(\bar{\bf Y}_{2}^{[T]}\mid \bar{\bf X}_{1}^{[T]},\mathcal{G})- H(\bar{\bf Y}_{2}^{[T]}\mid \bar{\bf X}_{1}^{[T]},\bar{\bf X}_{2}^{[T]},\mathcal{G})\label{jl2}\\
TR_3&\le& H(\bar{\bf Y}_{3}^{[T]}\mid  \bar{\bf X}_{1}^{[T]},\bar{\bf X}_{2}^{[T]},\mathcal{G})\label{jl3}
\end{eqnarray}
Therefore, for $\alpha\le1$, from \eqref{jl0}-\eqref{jl3} we have,
\begin{eqnarray}
TR_1+TR_2+TR_3&\le& H(\bar{\bf Y}_{1}^{[T]}\mid \mathcal{G})+4T(1-\alpha)\log{\bar{P}}+T~o~(\log{\bar{P}})\\
&\le&T(6-3\alpha)\log{\bar{P}}+T~o~(\log{\bar{P}})\label{oi}
\end{eqnarray}
\eqref{oi} is true as discrete entropy of any discrete random variable is bounded by logarithm of its cardinality.

\subsection{Equivalent Bounds}
Theorem \ref{theorem:GDoF1} is concluded from the following bounds,
\begin{enumerate}
\item If $\alpha\in\mathbb{R}^+,\alpha\le\frac{1}{2}$, then 
\begin{eqnarray}
&&\sum_{k=1}^Kd_k\nonumber\\
&\le& \frac{K\left(\min(M,N)(1-\alpha)+(N-M)^+\alpha\right)+K(K-2)\min(M,N)(1-\alpha)}{K-1}\label{b1}
\end{eqnarray}
\item If $\alpha\in\mathbb{R}^+,\frac{1}{2}\le\alpha\le1$, then 
\begin{eqnarray}
&&\sum_{k=1}^Kd_k\nonumber\\
&\le& \frac{K\left(\min((K-1)M,N)\alpha+(N-(K-1)M)^+(1-\alpha)\right)+K(K-2)\min(M,N)(1-\alpha)}{K-1}\nonumber\\
&&\label{b1+}
\end{eqnarray}
\item If $\alpha\in\mathbb{R}^+,\alpha\le1$, then 
\begin{eqnarray}
\sum_{k=1}^Kd_k&\le& N\alpha+K\min(M,N)(1-\alpha)\label{b2}
\end{eqnarray}
\item If $\alpha\in\mathbb{R}^+,1\le\alpha$, then 
\begin{eqnarray}
\sum_{k=1}^Kd_k&\le& (N-(K-1)M)^++\min(N,(K-1)M)\alpha\label{b2+}
\end{eqnarray}
\item For any $\alpha\in \mathbb{R}^+$, 
\begin{eqnarray}
\sum_{k=1}^Kd_k&\le& K\min(M,N)\label{b3}
\end{eqnarray}
\end{enumerate}
Thus, in order to prove Theorem \ref{theorem:GDoF1}, the Bounds \eqref{b1}-\eqref{b3} should be proved.
\subsection{Proof of Bounds \eqref{b1}-\eqref{b3}}\label{mt}
The last bound,  $\sum_{k=1}^Kd_k\le K\min(M,N)$ is the trivial combination of single user bounds.  Let us prove the other four  bounds, i.e., \eqref{b1}-\eqref{b2+}. 
\begin{enumerate}
\item {\bf Proof of \eqref{b1} and \eqref{b1+}}\\

 Writing Fano's Inequality for the first $K-1$ receivers we have,
\begin{eqnarray}
TR_1&\le& I(\bar{\bf Y}^{[T]}_{1};\bar{\bf X}^{[T]}_{1}\mid \mathcal{G})\label{c1}\\
TR_k&\le& I(\bar{\bf Y}^{[T]}_{k};\bar{\bf X}^{[T]}_{k}\mid \bar{\bf X}^{[T]}_{1},\cdots,\bar{\bf X}^{[T]}_{k-1},\mathcal{G}), \forall k\in[K-1],k\neq 1\label{c2}
\end{eqnarray}
{Summing}  \eqref{c1} and \eqref{c2}, we have,
\begin{eqnarray}
&&T\sum_{k=1}^{K-1}R_k\nonumber\\
&\le& I(\bar{\bf Y}^{[T]}_{1};\bar{\bf X}^{[T]}_{1}\mid \mathcal{G})+\sum_{k=2}^{K-1} I(\bar{\bf Y}^{[T]}_{k};\bar{\bf X}^{[T]}_{k}\mid \bar{\bf X}^{[T]}_{1},\cdots,\bar{\bf X}^{[T]}_{k-1},\mathcal{G})\\
&=& H(\bar{\bf Y}^{[T]}_{1}\mid \mathcal{G})-H\left(\bar{\bf X'}^{[T]}_{K}\mid \mathcal{G}\right)\nonumber\\
&&+\sum_{k=2}^{K-1} \left(H(\bar{\bf Y}^{[T]}_{k}\mid  \bar{\bf X}^{[T]}_{1},\cdots,\bar{\bf X}^{[T]}_{k-1},\mathcal{G})-H(\bar{\bf Y}^{[T]}_{k-1}\mid  \bar{\bf X}^{[T]}_{1},\cdots,\bar{\bf X}^{[T]}_{k-1},\mathcal{G})\right)\\
&\le& H(\bar{\bf Y}^{[T]}_{1}\mid \mathcal{G})-H\left(\bar{\bf X'}^{[T]}_{K}\mid \mathcal{G}\right)+\sum_{k=2}^{K-1} T\min(M,N)(1-\alpha)\log{\bar{P}}+T~o(\log{\bar{P}})\label{c3}
\end{eqnarray}
where the new random variable,  $\bar{\bf X'}_k(t)$ is defined as\\
\begin{eqnarray}
\bar{\bf X'}_k(t)&=&\begin{bmatrix}\bar{X'}_{k1}(t)&\bar{X'}_{k2}(t)&\cdots&\bar{X'}_{kN}(t)\end{bmatrix}^\dagger\\
\bar{X'}_{kn}(t)&=&{L}_{kn3}^b(t)\left((\bar{\bf X}_{k}(t))^{\alpha}\right), \forall n\in[N]
\end{eqnarray}
Let us explain how Lemma \ref{lemma} {yields}  \eqref{c3}. Substitute the random variables $\bar{\bf U}_1$, $\bar{\bf U}_2$, $\bar{\bf V}_1^{[T]}$, $\bar{\bf V}_2^{[T]}$ and $W$  in Lemma \ref{lemma}  with $\bar{\bf Y}^{[T]}_{k}$, $\bar{\bf Y}^{[T]}_{k-1}$, $\bar{\bf X}_k^{[T]}$,  $\left(\bar{\bf X}_{j}^{[T]},j\in[K],j\notin[k]\right)$ and $\left(\bar{\bf X}_{j}^{[T]},j\in[k-1]\right)$, respectively. Next,  we set $\eta=1, \lambda_{11}=1, \lambda_{21}=\alpha, \lambda_{12}=\alpha, \lambda_{22}=\alpha,M_1=M,M_2=(K-k)M,N_1=N_2=N$. Thus, we have $(\lambda_{11}-\lambda_{21})^+=1-\alpha$ and $(\lambda_{12}-\lambda_{22})^+=0$. Therefore, from Lemma \ref{lemma},  \eqref{c3} is concluded. Similar to  \eqref{c3}, by symmetry we have, 
\begin{eqnarray}
&&T\sum_{k\in[K],k\neq j}R_k\nonumber\\
&\le& H(\bar{\bf Y}^{[T]}_{j+1}\mid \mathcal{G})-H\left(\bar{\bf X'}^{[T]}_{j}\mid \mathcal{G}\right)\nonumber\\
&&+(K-2) T\min(M,N)(1-\alpha)\log{\bar{P}}+T~o(\log{\bar{P}})\label{c4}
\end{eqnarray}
for all $j\in[K]$. Summing \eqref{c4} for all $j\in[K]$ we have,
\begin{eqnarray}
&&T(K-1)\sum_{k=1}^KR_k\nonumber\\
&=&T\sum_{j=1}^K\sum_{k\in[K],k\neq j}R_k\nonumber\\
&\le& \sum_{k=1}^K\left(H(\bar{\bf Y}^{[T]}_{k}\mid \mathcal{G})-H\left(\bar{\bf X'}^{[T]}_{k}\mid \mathcal{G}\right)\right)+TK(K-2) \min(M,N)(1-\alpha)\log{\bar{P}}\nonumber\\
&&+T~o(\log{\bar{P}})\label{xx}
\end{eqnarray}
Now, let us consider the two cases of $\alpha\le\frac{1}{2}$ and $\frac{1}{2}\le\alpha\le1$ separately.
\begin{enumerate}
\item{$\alpha\le\frac{1}{2}$}
\begin{eqnarray}
&&H(\bar{\bf Y}^{[T]}_{k}\mid \mathcal{G})-H\left(\bar{\bf X'}^{[T]}_{k}\mid \mathcal{G}\right)\nonumber\\
&\le&  T\left(\min(M,N)(1-\alpha)+(N-M)^+\alpha\right)\log{\bar{P}}+T~o(\log{\bar{P}})\label{c5}
\end{eqnarray}
Let us explain how \eqref{c5} follows from Lemma \ref{lemma}. Substitute the random variables  $\bar{\bf U}_1$, $\bar{\bf U}_2$, $\bar{\bf V}_1^{[T]}$ and $\bar{\bf V}_2^{[T]}$  with $\bar{\bf Y}^{[T]}_{k}$, $\bar{\bf X'}^{[T]}_{k}$, $\bar{\bf X}_k^{[T]}$ and $\left(\bar{\bf X}_{j}^{[T]},j\in[K],j\neq k\right)$, respectively. Moreover,  setting $\eta=1, \lambda_{11}=1, \lambda_{21}=\alpha, \lambda_{12}=\alpha, \lambda_{22}=0$, we have $(\lambda_{11}-\lambda_{21})^+=1-\alpha$ and $(\lambda_{12}-\lambda_{22})^+=\alpha$. Therefore, from Lemma \ref{lemma} we conclude \eqref{c5}. From \eqref{xx} and \eqref{c5}, \eqref{b1} is concluded.
\item{$\frac{1}{2}\le\alpha\le1$}
\begin{eqnarray}
&&H(\bar{\bf Y}^{[T]}_{k}\mid \mathcal{G})-H\left(\bar{\bf X'}^{[T]}_{k}\mid \mathcal{G}\right)\nonumber\\
&\le&  T\left(\min((K-1)M,N)\alpha+(N-(K-1)M)^+(1-\alpha)\right)\log{\bar{P}}+T~o(\log{\bar{P}})\label{c5g}
\end{eqnarray}
 \eqref{c5g} follows from Lemma \ref{lemma} similar to \eqref{c5}. Substitute the random variables  $\bar{\bf U}_1$, $\bar{\bf U}_2$, $\bar{\bf V}_1^{[T]}$ and $\bar{\bf V}_2^{[T]}$  with $\bar{\bf Y}^{[T]}_{k}$, $\bar{\bf X'}^{[T]}_{k}$, $\left(\bar{\bf X}_{j}^{[T]},j\in[K],j\neq k\right)$ and $\bar{\bf X}_k^{[T]}$, respectively. The rest of the proof is concluded similar to \eqref{c5}. From \eqref{xx} and \eqref{c5g}, \eqref{b1+} is concluded.
\end{enumerate}
\item {\bf Proof of \eqref{b2}}\\

 Summing \eqref{c1} and \eqref{c2}, we have,
\begin{eqnarray}
&&T\sum_{k=1}^{K}R_k\nonumber\\
&\le& I(\bar{\bf Y}^{[T]}_{1};\bar{\bf X}^{[T]}_{1}\mid \mathcal{G})+\sum_{k=2}^{K} I(\bar{\bf Y}^{[T]}_{k};\bar{\bf X}^{[T]}_{k}\mid \bar{\bf X}^{[T]}_{1},\cdots,\bar{\bf X}^{[T]}_{k-1},\mathcal{G})\label{c6+}\\
&=& H(\bar{\bf Y}^{[T]}_{1}\mid \mathcal{G})\nonumber\\
&&+\sum_{k=2}^{K} \left(H(\bar{\bf Y}^{[T]}_{k}\mid  \bar{\bf X}^{[T]}_{1},\cdots,\bar{\bf X}^{[T]}_{k-1},\mathcal{G})-H(\bar{\bf Y}^{[T]}_{k-1}\mid  \bar{\bf X}^{[T]}_{1},\cdots,\bar{\bf X}^{[T]}_{k-1},\mathcal{G})\right)\\
&\le& H(\bar{\bf Y}^{[T]}_{1}\mid \mathcal{G})+\sum_{k=2}^{K} T\min(M,N)(1-\alpha)\log{\bar{P}}+T~o(\log{\bar{P}})\label{c6}\\
&\le& T\left((N-M)^+\alpha+\min(M,N)\right)\log{\bar{P}}+\sum_{k=2}^{K} T\min(M,N)(1-\alpha)\log{\bar{P}}\nonumber\\
&&+T~o(\log{\bar{P}})\label{cv6}\\
&=& T(N\alpha+K\min(M,N)(1-\alpha))\log{\bar{P}}+T~o(\log{\bar{P}})\label{c7}
 \end{eqnarray}
\eqref{c6} follows similar to \eqref{c3} and \eqref{cv6}  is concluded as the entropy of a discrete random variable is   bounded by logarithm of the cardinality of its support, i.e., \footnote{\eqref{cv6} follows from Lemma \ref{lemma}  by substituting $\bar{\bf U}_1$ and $\bar{\bf U}_2$ with $\bar{\bf Y}^{[T]}_{1}$ and ${ C}^{[T]}$ where ${\bf C}^{[T]}$ is a $T$-letter constant variable.  Then, substituting $\bar{\bf V}_1^{[T]}$ and $\bar{\bf V}_2^{[T]}$  with  $\bar{\bf X}_1^{[T]}$ and $\left(\bar{\bf X}_{j}^{[T]},j\in[K],j\neq 1\right)$, \eqref{cv6} is concluded. Here, we assume $\eta=1, \lambda_{11}=1, \lambda_{21}=\alpha, \lambda_{12}=0, \lambda_{22}=0,,M_1=M,M_2=(K-1)M,N_1=N_2=N$.}
\begin{eqnarray}
 H(\bar{\bf Y}^{[T]}_{1}\mid \mathcal{G})&\le& T\left((N-M)^+\alpha+\min(M,N)\right)\log{\bar{P}}
 \end{eqnarray}
Dividing \eqref{c7} by $T\log{\bar{P}}$, \eqref{b2}  is obtained.
\item {\bf Proof of \eqref{b2+}}\\

 Similarly, from (\eqref{c6+}-\eqref{c6}) we have,
\begin{eqnarray}
&&T\sum_{k=1}^{K}R_k\nonumber\\
&\le& H(\bar{\bf Y}^{[T]}_{1}\mid \mathcal{G})+\sum_{k=2}^{K} T\min(M,N)(1-\alpha)^+\log{\bar{P}}+T~o(\log{\bar{P}})\\
&\le& H(\bar{\bf Y}^{[T]}_{1}\mid \mathcal{G})+T~o(\log{\bar{P}})\label{c6-}\\
&\le& (N-(K-1)M)^+T\log{\bar{P}}+\min(N,(K-1)M)\alpha T\log{\bar{P}}+T~o(\log{\bar{P}})\label{c6--}
 \end{eqnarray}
\eqref{c6-} is true as $1\le\alpha$ and \eqref{c6--} follows similar\footnote{\eqref{c6--} follows similar to  \eqref{cv6}  from Lemma \ref{lemma}. Substitute $\bar{\bf U}_1$, $\bar{\bf U}_2$, $\bar{\bf V}_1^{[T]}$ and $\bar{\bf V}_2^{[T]}$ with $\bar{\bf Y}^{[T]}_{1}$, ${ C}^{[T]}$,  $\left(\bar{\bf X}_{j}^{[T]},j\in[K],j\neq 1\right)$ and $\bar{\bf X}_1^{[T]}$ and  assume $\eta=\alpha, \lambda_{11}=\alpha, \lambda_{21}=1, \lambda_{12}=0, \lambda_{22}=0,M_1=(K-1)M,M_2=M,N_1=N_2=N$.} to  \eqref{cv6}. Dividing \eqref{c6--} by $T\log{\bar{P}}$, \eqref{b2+}  is obtained.

\end{enumerate}

\section{Proof of Theorem \ref{theorem:GDoF1}: Achievability}
\subsection{A Useful Lemma}
Consider a $(M_1+M_2)$-user multiple access channel (MAC) where each transmitter is equipped with a single antenna, the receiver has $N$ antennas, $N< M_1+M_2$, and the $N\times 1$ received signal vector ${\bf Q}$ is represented as,
\begin{align}
{\bf Q}=&\sqrt{P}\sum_{k=1}^{M_1} {\bf H}_k{ T}_k+\sqrt{P^{\alpha}}\sum_{k=M_1+1}^{M_1+M_2} {\bf H}_k{ T}_k+\sum_{n=1}^{N} \sqrt{P^{\alpha_n}}{\bf G}_nZ_n\label{mac0}
\end{align}
where  $T_1, T_2, \cdots, T_{M_1+M_2}$ are the transmitted signals, and $Z_1,Z_2,\cdots,Z_N$ are i.i.d. Gaussian zero mean unit variance noise terms. The ${\bf H}_k, {\bf G}_n$ are $N\times 1$ generic vectors, i.e.,  generated from continuous distributions with bounded density, so that any $N$ of them are linearly independent almost surely. The transmit power constraint is  expressed as,
\begin {eqnarray}
\mbox{E}{|T_{k}|}^2&\leq&P^{-\eta_k},~\forall k\in[M_1+M_2]\label{mac1}
\end{eqnarray}
where for any $k\in[M_1+M_2]$,  $\eta_k$ is a non-negative integer. Further, define $\gamma_k$ for $k\in[M_1+M_2]$ as,
\begin {eqnarray}
\gamma_k&=&\left\{
\begin{array}{ll} 
{(1-\eta_k)}^+,&k\in[M_1]\\ 
{(\alpha-\eta_k)}^+, &\text{Otherwise}
\end{array}
\right.\label{gamma1}
\end{eqnarray}
Thus $\gamma_k$ is the received power level of user $k$ in the GDoF sense. The GDoF region $\mathcal{D}'$ is defined as
\begin{align}
\mathcal{D}'\triangleq&\{(d'_1,d'_2,\cdots,d'_{M_1+M_2}): \exists (R'_1(P),R'_2(P),\cdots,R'_{M_1+M_2}(P))\in\mathcal{C}'(P),\nonumber\\
& \mbox{ s.t. } d'_k=\lim_{P\rightarrow\infty}\frac{R'_k(P)}{\frac{1}{2}\log{(P)}}, \forall k\in[M_1+M_2]\} \label {region}
\end{align}
 where $\mathcal{C}'(P)$ is the capacity region of MAC described in (\ref{mac0}).

\begin{lemma}\label{lemma:mac} The GDoF tuple $(d'_1, d'_2, \cdots, d'_{M_1+M_2})$ is achievable in the multiple access channel described above if $\forall k\in[M_1+M_2]$, and $\forall S\subset[M_1+M_2]$ where $|S|=k$,
\begin{align}
\sum_{i\in S} d'_i
&\le\max_{S_2\in S,|S_2|=\min(k,N)}\sum_{i\in S_2} \gamma_i -\min_{S_1\in[N],|S_1|=\min(k,N)}\sum_{i\in S_1} \alpha_i \label{mac3}
\end{align}
\end{lemma} 
For proof of Lemma \ref{lemma:mac} see \cite{Arash_Jafar_MIMOsym_ArXiv}. It is sufficient to derive the achievability for Theorem \ref{theorem:GDoF1}, as Theorem \ref{theorem:GDoF1} is automatically concluded from it.

\subsection{Proof of Achievability in Theorem \ref{theorem:GDoF1}} \label{app3}
Now, let us achieve the bound (\ref{fg<}). We will suppress the time-index $t$ in this section to simplify the notation. For any $k\in[K]$ user $k$'s message $W_k$ is split into messages $(W_{kc},W_{kp})$, representing common message and private message, respectively.  Let us consider the three cases of $\alpha\le\frac{1}{2}$, $\frac{1}{2}\le\alpha\le1$, and $1\le\alpha$ separately as,
\begin{enumerate}
\item {$\alpha\le\frac{1}{2}$.} Our goal here is to achieve $\min(M,N)(1-\alpha)+\frac{(N-M)^+\alpha}{K-1}$ GDoF per user where results in $K\min(M,N)(1-\alpha)+\frac{K(N-M)^+\alpha}{K-1}$ GDoF totally. In order to achieve $\min(M,N)(1-\alpha)+\frac{(N-M)^+\alpha}{K-1}$ GDoF per user, for any $k\in[K]$ the public message $W_{kc}$ is encoded into Gaussian codebooks $U_{k1c},U_{k2c},\cdots,U_{kMc}$ with powers $\mbox{E}{|U_k|}^2=1-P^{-\alpha}$ each carrying $\frac{(N-M)^+\alpha}{(K-1)M}$ GDoF. These codewords are transmitted through $M$ antennas along  $M\times 1$ generic unit  vectors ${\bf V}_{k1},{\bf V}_{k2},\cdots,{\bf V}_{k{M}}$.  The private message $W_{kp}$ is encoded into Gaussian codebooks $U_{k1p},U_{k2p},\cdots,U_{k\min(M,N)p}$ with powers $\mbox{E}{|U_{kjp}|}^2=P^{-\alpha}$ for any $j\in[\min(M,N)]$ so that the total power per transmitter is unity. These codewords are transmitted through $\min(M,N)$ antennas along the $M\times 1$ generic unit  vectors ${\bf V}_{k1},{\bf V}_{k2},\cdots,{\bf V}_{k{\min(M,N)}}$. Each of the  private messages is carrying $1-\alpha$ GDoF. The transmitted and received signals are,
\begin{align}
\mathbf{X}_{k}=&\sum_{j=1}^{M}\mathbf{V}_{kj}{U}_{kjc}+\sum_{j=1}^{\min{M,N}}\mathbf{V}_{kj}{U}_{kjp}\label{TR21a-}\\
\mathbf{Y}_{k}=&\sqrt{P}{\bf G}_{kk}\mathbf{X}_{k}+\sum_{j=1,j\neq k}^{K}\sqrt{P^{\alpha}}{\bf {G}}_{kj}\mathbf{X}_{j}+\mathbf{\Gamma}_{k}\label{TR21a}
\end{align}
 Using Lemma \ref{lemma:mac} we claim that each receiver, e.g., receiver $1$ can decode  all the signals $U_{kjc}$ and $U_{1jp}$ for all $k\in[K]$ and $j\in[M]$  treating all the other signals as noise. Set the variables $M_1=M+\min(M,N)$, $M_2=(K-1)M$ and $\alpha_n=0$ for all $n\in[N]$. Moreover, define the codewords $T_{1},\cdots,T_{KM+\min(M,N)}$ as 
\begin {align}
T_j=&\left\{
\begin{array}{ll} 
U_{1jc},&1\le j\le {M}\\ 
U_{2(j-M)c},&M<j\le 2M\\
\vdots &\vdots\\
U_{K(j-(K-1)M)c},&(K-1)M<j\le KM\\
U_{1(j-KM)p},&KM<j\le KM+\min(M,N)
\end{array}
\right.
\end{align}
From \eqref{gamma1}, $\gamma_1,\cdots,\gamma_{KM+\min(M,N)}$ are derived as,
\begin {align}
\gamma_j=&\left\{
\begin{array}{ll} 
1,&1\le j\le M\\ 
\alpha,&M< j\le KM\\ 
1-\alpha,&KM< j\le KM+\min(M,N)
\end{array}
\right.\label{ffd2}
\end{align}
Note that $N\le KM$ and $\alpha\le\frac{1}{2}$. Thus, from the received signal in (\ref{TR21a}), $T_1,\cdots,T_{KM+\min(M,N)}$ are decoded by  first receiver as (\ref{mac3}) is satisfied for all $k\in[KM+\min(M,N)]$. For instance if we set $k=KM+\min(M,N)$, the condition (\ref{mac3}) is equivalent to,

\begin{align}
\sum_{i\in S} d'_i=&\frac{K(N-M)^+\alpha}{K-1}+\min(M,N)(1-\alpha)\nonumber\\
\le&\min(M,N)+(N-M)^+\alpha=\max_{S_2\in S,|S_2|=\min(k,N)}\sum_{i\in S_2} \gamma_i .
\end{align}

\item{$\frac{1}{2}<\alpha\leq1$.} Let us achieve $d$ GDoF where $d$  is equal to,
\begin{align}
d=\min\left(\frac{K}{K-1}\left((K-2)\min(M,N)(1-\alpha)+N(\alpha)\right),N\alpha+K\min(M,N)\left(1-\alpha\right)\right) 
\end{align}
Similar to the case $\alpha\le\frac{1}{2}$, the public message $W_{kc}$ is encoded into Gaussian codebooks $U_{k1c},U_{k2c},\cdots,U_{kMc}$ with powers $\mbox{E}{|U_k|}^2=1-P^{-\alpha}$ each carrying $\frac{d-K\min(M,N)(1-\alpha)}{KM}$ GDoF. These codewords are transmitted through $M$ antennas along  $M\times 1$ generic unit  vectors ${\bf V}_{k1},{\bf V}_{k2},\cdots,{\bf V}_{k{M}}$.  The private ~message $W_{kp}$ ~is encoded~ into Gaussian~ codebooks ~$U_{k1p},U_{k2p},\cdots,U_{k\min(M,N)p}$ with powers $\mbox{E}{|U_{kjp}|}^2=P^{-\alpha}$ for any $j\in[\min(M,N)]$. These codewords are transmitted through $\min(M,N)$ antennas along the $M\times 1$ generic unit  vectors ${\bf V}_{k1},{\bf V}_{k2},\cdots,{\bf V}_{k{\min(M,N)}}$. Each of the  private messages is carrying $1-\alpha$ GDoF. The transmitted and received signals follows the same as \eqref{TR21a-} and \eqref{TR21a}. From Lemma \ref{lemma:mac} each receiver, e.g., receiver $1$ can decode all the codewords $U_{kjc}$ and $U_{1jn}$ for all $k\in[K]$ and $j\in[M]$  treating all the other signals as noise. The details how receiver $1$ can decode all these codewords follows the same as the case $\alpha\le\frac{1}{2}$.

\item{$1\le\alpha$.} In this case, $\min\left(D(\alpha),K\min(M,N)\right)$ is achieved as follows. Recall that $D(\alpha)$ was defined as $(N-(K-1)M)^++\min(N,(K-1)M)\alpha$. All $\min(M,N)$ messages of each transmitter are
public in this case and are encoded into Gaussian codebooks
$U_{k1c},U_{k2c},\cdots,U_{k\min(M,N)c}$ with unit powers each carrying $\min\left(\frac{D(\alpha)}{K\min(M,N)},1\right)$ GDoF. These codewords are transmitted through $\min(M,N)$ antennas along the $M\times 1$ generic unit  vectors ${\bf V}_{k1},{\bf V}_{k2},\cdots,{\bf V}_{k{\min(M,N)}}$. The transmitted and received signals are concluded similar to \eqref{TR21a-} and \eqref{TR21a} as,
\begin{align}
\mathbf{X}_{k}=&\sum_{j=1}^{M}\mathbf{V}_{kj}{U}_{kjc}\label{TR21a-1}\\
\mathbf{Y}_{k}=&\sqrt{P}{\bf G}_{kk}\mathbf{X}_{k}+\sum_{j=1,j\neq k}^{K}\sqrt{P^{\alpha}}{\bf {G}}_{kj}\mathbf{X}_{j}+\mathbf{\Gamma}_{k}\label{TR21a1}
\end{align}
Each receiver, e.g., receiver $1$ can decode  all the signals $U_{kjc}$ for all $k\in[K]$ and $j\in[\min(M,N)]$  treating all the other signals as noise. Set the variables $M_1=\min(M,N)$, $M_2=(K-1)\min(M,N)$ and $\alpha_n=0$ for all $n\in[N]$ in Lemma \ref{lemma:mac} . Moreover, define the codewords $T_{1},\cdots,T_{K\min(M,N)}$ as 
\begin {align}
T_j=&\left\{
\begin{array}{ll} 
U_{1jc},&1\le j\le {\min(M,N)}\\ 
U_{2(j-\min(M,N))c},&\min(M,N)<j\le 2\min(M,N)\\
\vdots &\vdots\\
U_{K(j-(K-1)\min(M,N))c},&(K-1)\min(M,N)<j\le K\min(M,N)
\end{array}
\right.
\end{align}
From \eqref{gamma1}, $\gamma_1,\cdots,\gamma_{K\min(M,N)}$ are derived as,
\begin {align}
\gamma_j=&\left\{
\begin{array}{ll} 
1,&1\le j\le \min(M,N)\\ 
\alpha,&\min(M,N)< j\le K\min(M,N)
\end{array}
\right.\label{ffd2}
\end{align}
Similar to the case $\alpha\le \frac{1}{2}$, $T_1,\cdots,T_{K\min(M,N)}$ are decoded by  first receiver as (\ref{mac3}) is satisfied for all $k\in[K\min(M,N)]$. For instance if we set $k=K\min(M,N)$, the condition (\ref{mac3}) is equivalent to,
\begin{align}
\sum_{i\in S} d'_i=&K\min(M,N)\times\min\left(\frac{D(\alpha)}{K\min(M,N)},1\right)\nonumber\\
\le&\min\left((K-1)M,N\right)\alpha+\left(N-(K-1)M\right)^+=\max_{S_2\in S,|S_2|=\min(k,N)}\sum_{i\in S_2} \gamma_i 
\end{align}

\end{enumerate}

\section{Conclusion}
Symmetric $K$-user MIMO IC with $M$ antennas at each transmitter and $N$ antennas at each receiver is considered. Sum GDoF of this channel is derived. The Sum GDoF  is found to be a $W$ curve as a function of $\alpha$ for fixed  $M$ and $N$ similar to the SISO case. Outer bound proof is obtained with the help of a key lemma that generalizes the AIS argument.  The achievability follows from the achievability of the GDoF region of a MAC, combined with the `treating interference as noise' scheme. 

\appendix

\section{Proof of Lemma \ref{lemma}}\label{lemmap}
 Define the random variables $[\bar{\bf U}_2]_{N_1}$ as,
\begin{eqnarray}
[\bar{\bf U}_2]_{N_1}&=&\left({U}_{21}^{[T]},{U}_{22}^{[T]},\cdots,{U}_{2N_1}^{[T]}\right)\label{lemmamimox2+}
\end{eqnarray}
As $H([{\bar{\bf U}}_2]_{N_1}\mid {W},\mathcal{G})\le H({\bar{\bf U}}_2\mid {W},\mathcal{G})$, it is sufficient to prove the inequality \eqref{lemmamimox5} for $N_1=N_2$. So from now on, we assume $N_1=N_2$. Before proceeding to prove \eqref{lemmamimox5}, note that for any $e\times1$ vector discrete random variable ${\bf V}$ and $e\times e$ matrix $A$,
\begin{eqnarray}
 H({\bf V})=H(A{\bf V})\text{~~if~~} |A|\neq0.
\end{eqnarray}
As multiplying invertible matrix to the vector discrete random variable does not change the entropy of it, it is sufficient to prove \eqref{lemmamimox5} for the random variables $\breve{\bf U}_{1}$ and $\breve{\bf U}_{2}$ which are defined as,
\begin{eqnarray}
\breve{\bf U}_1&=&\left(\breve{U}_{11}^{[T]},\breve{U}_{12}^{[T]},\cdots,\breve{U}_{1N_1}^{[T]}\right)\label{lemmamimox7}\\
\breve{\bf U}_2&=&\left(\breve{U}_{21}^{[T]},\breve{U}_{22}^{[T]},\cdots,\breve{U}_{2N_1}^{[T]}\right)\label{lemmamimox8}
\end{eqnarray}
where for any $i\in[2],t\in[T]$, $\breve{U}_{in}(t)$ are defined as,
\begin{eqnarray}
\breve{U}_{in}(t)&=&\left\{\begin{matrix}
 L_{in1}^b(t)\left((\bar{\mathbf{V}}_1(t))^{\eta}_{\eta-\lambda_{i1}}\bigtriangledown\cdots\bigtriangledown(\bar{\mathbf{V}}_l(t))^{\eta}_{\eta-\lambda_{il}}\right),~~&\forall n\in[M_1]\\ 
 L_{in2}^b(t)\left((\bar{\mathbf{V}}_2(t))^{\eta}_{\eta-\lambda_{i2}}\bigtriangledown\cdots\bigtriangledown(\bar{\mathbf{V}}_l(t))^{\eta}_{\eta-\lambda_{il}}\right),~& \forall n\in[M_1+M_2],n\notin[M_1]\\ 
\vdots& \vdots\\
L_{ins}^b(t)\left((\bar{\mathbf{V}}_s(t))^{\eta}_{\eta-\lambda_{is}}\bigtriangledown\cdots\bigtriangledown(\bar{\mathbf{V}}_l(t))^{\eta}_{\eta-\lambda_{il}}\right),&\forall n\in[\sum_{i=1}^sM_i],n\notin[\sum_{i=1}^{s-1}M_i]\\
 L_{in(s+1)}^b(t)\left((\bar{\mathbf{V}}_{s+1}(t))^{\eta}_{\eta-\lambda_{i(s+1)}} \bigtriangledown\right.&\\
\left.\cdots\bigtriangledown(\bar{\mathbf{V}}_l(t))^{\eta}_{\eta-\lambda_{il}}\right),&\forall n\in[N_1],n\notin[\sum_{i=1}^sM_i]
\end{matrix}\right. \label{lemmamimox9}
\end{eqnarray}
Thus, we have,
\begin{eqnarray}
&&H({\breve{\bf U}}_2\mid W,\mathcal{G})-H({\breve{\bf U}}_1\mid W,\mathcal{G})\nonumber\\
&=&H(\{\breve{U}_{2i}^{[T]},i\in[N_1]\}\mid W,\mathcal{G})-H(\{\breve{U}_{1i}^{[T]},i\in[N_1]\}\mid W,\mathcal{G})\label{sw1}\\
&=&\sum_{n=1}^{N_1}\big(H(\{\breve{U}_{1i'}^{[T]},\breve{U}_{2i}^{[T]},~\forall i,i'\in[N_1],i'<n\le i\}\mid W,\mathcal{G})\nonumber\\
&&-H(\{\breve{U}_{1i'}^{[T]},\breve{U}_{2i}^{[T]},~\forall i,i'\in[N_1],i'\le n<i\}\mid W,\mathcal{G})\big)\label{sw2}\\
&=&\sum_{n=1}^{N_1}\big(H(\breve{U}_{1n}^{[T]}\mid W,W_n,\mathcal{G})-H(\breve{U}_{2n}^{[T]}\mid W,W_n,\mathcal{G})\big)\label{sw3}\\
&\le&T\big((N_1-\sum_{i=1}^sM_i)(\lambda_{1,s+1}-\lambda_{2,s+1})^++\sum_{i=1}^sM_i(\lambda_{1i}-\lambda_{2i})^+\big)\log{\bar{P}}+T~o~(\log{\bar{P}})\label{lemmamimox5+}
\end{eqnarray}
where $W_n$ is defined as the set of random variables $\{\breve{U}_{1i'}^{[T]},\breve{U}_{2i}^{[T]},i,i'\in[N_1],i'<n<i\}$. \eqref{sw2} follows from definition of $\breve{U}_{in}(t)$ and \eqref{sw3} is a result of chain rule. \eqref{lemmamimox5+} is true as for any $n\in[N_1]$ we have,
\begin{eqnarray}
&&H(\breve{U}_{1n}^{[T]}\mid W,W_n,\mathcal{G})-H(\breve{U}_{2n}^{[T]}\mid W,W_n,\mathcal{G})\nonumber\\
&\le&\left\{\begin{matrix}
T(\lambda_{11}-\lambda_{21})^+\log{\bar{P}} +T~o~(\log{\bar{P}})&0<n\le M_1 \\ 
T(\lambda_{12}-\lambda_{22})^+\log{\bar{P}} +T~o~(\log{\bar{P}}) &M_1<n\le M_1+M_2 \\ 
\vdots&\vdots\\
T(\lambda_{1s}-\lambda_{2s})^+\log{\bar{P}} +T~o~(\log{\bar{P}}) &\sum_{i=1}^{s-1}M_i<n\le \sum_{i=1}^{s}M_i\\
T(\lambda_{1,s+1}-\lambda_{2,s+1})^+\log{\bar{P}} +T~o~(\log{\bar{P}})&\sum_{i=1}^{s}M_i<n\le N_1 
\end{matrix}\right.\label{lemma3final}
\end{eqnarray}
\subsection{Proof of \eqref{lemma3final}}
Without loss of generality, let us prove \eqref{lemma3final} for some arbitrary value of $n$, e.g., $n=1$. \eqref{lemma3final} follows for the other values of $n\in[N_1]$ similarly. We are interested in the maximum of $H(\breve{U}_{11}^{[T]}\mid \bar{W},\mathcal{G})-H(\breve{U}_{21}^{[T]}\mid \bar{W},\mathcal{G})$ over all possible random variables $\bar{\mathbf{V}}_i(t),i\in[l]$ where $\bar{W}$ is defined as $(W,W_1)$. Similar to the AIS approach in \cite{Arash_Jafar_PN}, we first claim that from the functional dependence argument without loss of generality the whole codeword $\breve{U}_{21}^{[T]}$ can be assumed as a function of $\breve{U}_{11}^{[T]},\bar{W},\mathcal{G}$. Therefore we have,
\begin {eqnarray}
&&H(\breve{U}_{21}^{[T]}\mid \bar{W},\mathcal{G})+H(\breve{U}_{11}^{[T]}\mid \breve{U}_{21}^{[T]},\bar{W},\mathcal{G})\nonumber\\
&=&H(\breve{U}_{11}^{[T]},\breve{U}_{21}^{[T]}\mid \bar{W},\mathcal{G})\label{ew1}\\
&=&H(\breve{U}_{11}^{[T]}\mid \bar{W},\mathcal{G})\label{ew2}
\end{eqnarray}
where $(\ref{ew1})$ follows from chain rule and $(\ref{ew2})$ is true as $\breve{U}_{21}^{[T]}$ is a function of $\breve{U}_{11}^{[T]},\bar{W},\mathcal{G}$. Thus, we should evaluate the maximum of $H(\breve{U}_{11}^{[T]}\mid \breve{U}_{21}^{[T]},\bar{W},\mathcal{G})$ for all possible random variables $\bar{\mathbf{V}}_i(t),i\in[l]$ where $\bar{\mathbf{V}}_i(t)=\begin{bmatrix}\bar{V}_{i1}(t)&\cdots&\bar{V}_{iM_i}(t)\end{bmatrix}^\dagger$ for any $i\in[l]$ and $\bar{V}_{im}(t)\in\mathcal{X}_{\eta}$ for any $i\in[l],m\in[M_i]$. In the next step, for a given $\bar{W}$ and channel realization $\mathcal{G}$, we define aligned image set $S_{\nu^{[T]}}(\bar{W},\mathcal{G})$ as the set of all values of $\breve{U}_{11}^{[T]}$ which produce  the same value for $\breve{U}_{21}^{[T]}$, as is produced by $\nu^{[T]}$. Since uniform distribution maximizes the entropy,
\begin{eqnarray}
\mathcal{D}_{\Delta}&\triangleq&H(\breve{U}_{11}^{[T]}\mid \bar{W},\mathcal{G})-H(\breve{U}_{21}^{[T]}\mid \bar{W},\mathcal{G})\nonumber\\
&\le& H(\breve{U}_{11}^{[T]}\mid \breve{U}_{21}^{[T]},\bar{W},\mathcal{G})\nonumber\\
&\le&\mbox{E}_{\mathcal{G}}\{\log{\left|S_{\nu^{[T]}}(\bar{W},\mathcal{G})\right| \}}\label{jen0}\\
&=&\mbox{E}_{\bar{W}}\left\{\mbox{E}_{\mathcal{G}}\{\log{|S_{\nu^{[T]}}(\bar{W},\mathcal{G})}|\mid \bar{W}\}\right\}\label{expw}\\
&\le&\max_{w\in\mathcal{W}}\mbox{E}_{\mathcal{G}}\{\log{|S_{\nu^{[T]}}(\bar{W},\mathcal{G})}|\mid \bar{W}=w\}\label{expw2}\\
&\le&\max_{w\in\mathcal{W}}\log{\left\{\mbox{E}_{\mathcal{G}}\{|S_{\nu^{[T]}}(\bar{W},\mathcal{G})|\mid \bar{W}=w\}\right\}}\label{jen}
\end{eqnarray}  
where $\mathcal{W}$ is support of the random variable $\bar{W}$. (\ref{jen}) is concluded from the Jensen's Inequality. Thus, in order to bound the maximum of $H(\breve{U}_{11}^{[T]}\mid \breve{U}_{21}^{[T]},\bar{W},\mathcal{G})$, we bound the cardinality of the aligned image set form above by using Bounded Density Assumption of $\mathcal{G}$. 

\subsubsection{Functional Dependence $\breve{U}_{21}^{[T]}(\breve{U}_{11}^{[T]},\bar{W},\mathcal{G})$} \label{funcdep}
Using the functional dependence argument as in \cite{Arash_Jafar_sumset}, henceforth we assume $\breve{U}_{21}^{[T]}$ is a function of $\breve{U}_{11}^{[T]},\bar{W},\mathcal{G}$.

\subsubsection { Definition of Aligned Image Sets} \label {defrt}
The aligned image set $S_{\nu^{[T]}}(\bar{W},\mathcal{G})$ for given $\bar{W}=w$ and realization $\mathcal{G}=G$  is defined as the non-empty set containing the codeword $\nu^{[T]}\in \mbox{supp}(\breve{U}_{11}^{[T]})$  and all the values   $\breve{U}_{11}^{[T]}$ that produces the same $\breve{U}_{21}^{[T]}$ value as is produced by $\breve{U}_{11}^{[T]}=\nu^{[T]}$. Mathematically,
\begin{eqnarray}
S_{\nu^{[T]}}({w},{G})&\triangleq&\{\bar{\nu}^{[T]}\in\mbox{supp}(\breve{U}_{11}^{[T]}): \breve{U}_{21}^{[T]}(\nu^{[T]},w,{G})=\breve{U}_{21}^{[T]}(\bar{\nu}^{[T]},w,{G})\}
\end{eqnarray}
As the cardinality $|S_{\nu^{[T]}}({w},{G})|$ as a function of $\mathcal{G}$, is a measurable function, we can compute the expected value of it \footnote{The cardinality $|S_{\nu^{[T]}}({w},{G})|$ as a function of $\mathcal{G}$, is a bounded simple function, and therefore measurable, see \cite{Arash_Jafar_PN}.}. It is bounded too as the members of the set are restricted to  the set of natural numbers not greater than $c\bar{P}^{\eta}$, where $c$  depends on $\Delta_2$. 

\subsubsection{Bounding the Probability of Image Alignment}
From (\ref{jen}),  $\mbox{E}_{\mathcal{G}}\{|S_{\nu^{[T]}}(\bar{W},\mathcal{G})|\mid \bar{W}=w\}=\sum_{\bar{\nu}^{[T]}\in\mbox{supp}(\breve{U}_{11}^{[T]})}P_a(\bar{\nu}^{[T]})$ should be computed. Given $\bar{W}=w$ and $\mathcal{G}$, consider two distinct realizations of $\breve{U}_{11}^{[T]}$, say $\bar{\nu}^{[T]}$, and $\breve{\nu}^{[T]}$, which are produced by two distinct  realizations of $\bar{\mathbf{V}}_1^{[T]},\cdots,\bar{\mathbf{V}}_l^{[T]}$, denoted as $\dot{\mathbf{V}}_1^{[T]},\cdots,\dot{\mathbf{V}}_l^{[T]}$ and $\ddot{\mathbf{V}}_1^{[T]},\cdots,\ddot{\mathbf{V}}_l^{[T]}$ where $\dot{\mathbf{V}}_i(t)=\begin{bmatrix}\mu_{i1}(t)&\cdots&\mu_{iM_i}(t)\end{bmatrix}^\dagger$, $\ddot{\mathbf{V}}_i(t)=\begin{bmatrix}\pi_{i1}(t)&\cdots&\pi_{iM_i}(t)\end{bmatrix}^\dagger$ for any $i\in[l]$ and $\mu_{im}(t),\pi_{im}(t)\in\mathcal{X}_{\eta}$ for any $i\in[l],m\in[M_i]$. 

We wish to bound the probability that the images of these two codewords align, or in other words 
$\breve{U}_{21}^{[T]}(\bar{\nu}^{[T]},\bar{W},\mathcal{G})=\breve{U}_{21}^{[T]}(\breve{\nu}^{[T]},\bar{W},\mathcal{G})$,
\begin{eqnarray}
 &&L^b(t)\left((\dot{\mathbf{V}}_1(t))^{\eta}_{\eta-\lambda_{21}}\bigtriangledown\cdots\bigtriangledown(\dot{\mathbf{V}}_l(t))^{\eta}_{\eta-\lambda_{2l}}\right)\nonumber\\
 &=&L^b(t)\left((\ddot{\mathbf{V}}_1(t))^{\eta}_{\eta-\lambda_{21}}\bigtriangledown\cdots\bigtriangledown(\ddot{\mathbf{V}}_l(t))^{\eta}_{\eta-\lambda_{2l}}\right), \forall t\in[T].\label{equal1}
\end{eqnarray}
We rewrite \eqref{equal1} as follows,
\begin{eqnarray}
\sum_{i=1}^l\sum_{j=1}^{M_i}\lfloor g_{ij} (t)(\mu_{ij}(t))_{\eta-\lambda_{2i}}^{\eta}\rfloor &=&
\sum_{i=1}^l\sum_{j=1}^{M_i}\lfloor g_{ij}(t) (\pi_{ij})(t)_{\eta-\lambda_{2i}}^{\eta}\rfloor\label{equal2}, \forall t\in[T].
\end{eqnarray}
As for any real number $x$ we have $|x-\lfloor x\rfloor|<1$, from \eqref{equal2}  we conclude that
\begin{eqnarray}
\left|\sum_{i=1}^l\sum_{j=1}^{M_i}g_{ij}(t) \left( (\mu_{ij}(t))_{\eta-\lambda_{2i}}^{\eta}- (\pi_{ij}(t))_{\eta-\lambda_{2i}}^{\eta}\right)\right|\le \sum_{i=1}^lM_i \label{equal3}
\end{eqnarray}
So for fixed values of $g_{ij}(t),(i,j)\neq (1,1)$ the random variable $g_{11}(t)  \left( (\mu_{11}(t))_{\eta-\lambda_{21}}^{\eta}- (\pi_{11}(t))_{\eta-\lambda_{21}}^{\eta}\right)$ must take values within an interval of length no more than $2\sum_{i=1}^lM_i$. If $(\mu_{11}(t))_{\eta-\lambda_{21}}^{\eta}\neq (\pi_{11}(t))_{\eta-\lambda_{21}}^{\eta}$, then $g_{11}(t)$ must take values in an interval of length no more than $\frac{2\sum_{i=1}^lM_i}{\left|(\mu_{11}(t))_{\eta-\lambda_{21}}^{\eta}- (\pi_{11}(t))_{\eta-\lambda_{21}}^{\eta}\right|}$, the probability of which is no more than $\frac{2\sum_{i=1}^lM_if_{\max}}{\left|(\mu_{11}(t))_{\eta-\lambda_{21}}^{\eta}- (\pi_{11}(t))_{\eta-\lambda_{21}}^{\eta}\right|}$ \footnote{Note that the integral of any real-valued measurable function $h(x)$ over any measurable set $S$ can be bounded above by $\max_{x\in \mathcal {R}}h(x)$ times the measure of the set $S$ \cite{stein}.}. This is true since $\bar{V}_{im}(t)$ and $\mathcal{G}$ are independent  for any $i\in[l],m\in[M_i]$. Similarly, instead of $g_{11}(t)$ consider $g_{rs}(t)$ for any $r\in[l],s\in[M_r]$. The probability of the inequality \eqref{equal3} will be bounded by $\frac{2\sum_{i=1}^lM_if_{\max}}{\left|(\mu_{rs}(t))_{\eta-\lambda_{2r}}^{\eta}- (\pi_{rs}(t))_{\eta-\lambda_{2r}}^{\eta}\right|}$ if $(\mu_{rs}(t))_{\eta-\lambda_{2r}}^{\eta}\neq (\pi_{rs}(t))_{\eta-\lambda_{2r}}^{\eta}$. Therefore, considering all the $T$ channel uses, the probability of alignment of $\bar{\nu}^{[T]}$ with $\breve{\nu}^{[T]}$ is bounded by,
\begin{eqnarray}
P_a(\bar{\nu}^{[T]})&\le&\prod_{t=1}^T\min\left(1,\frac{2\sum_{i=1}^lM_if_{\max}}{\max_{r\in[l],s\in[M_r]}\left|(\mu_{rs}(t))_{\eta-\lambda_{2r}}^{\eta}- (\pi_{rs}(t))_{\eta-\lambda_{2r}}^{\eta}\right|}\right)\label{u000}
\end{eqnarray}
However, we wish to express the probability of alignment in the terms of $\bar{\nu}^{[T]}$ and $\breve{\nu}^{[T]}$. From \eqref{lemmamimox9}, the codewords $\bar{\nu}(t)$ and $\breve{\nu}(t)$ can be expressed as,
\begin{eqnarray}
\bar{\nu}(t)&=&\sum_{i=1}^l\sum_{j=1}^{M_i}\lfloor g_{ij} (t)(\mu_{ij}(t))_{\eta-\lambda_{1i}}^{\eta}\rfloor, \forall t\in[T]\label{u0}\\
\breve{\nu}(t)&=&\sum_{i=1}^l\sum_{j=1}^{M_i}\lfloor g_{ij}(t) (\pi_{ij})(t)_{\eta-\lambda_{1i}}^{\eta}\rfloor\label{equal2}, \forall t\in[T].\label{u00}
\end{eqnarray}
Thus, we have,
\begin{eqnarray}
&&|\bar{\nu}(t)-\breve{\nu}(t)|\nonumber\\
&\le&\sum_{i=1}^lM_i+\sum_{i=1}^l\sum_{j=1}^{M_i}|g_{ij} (t)|\times|(\mu_{ij}(t))_{\eta-\lambda_{1i}}^{\eta}-(\pi_{ij}(t))_{\eta-\lambda_{1i}}^{\eta}|\label{u1}\\
&\le&\sum_{i=1}^lM_i+\Delta_2\left(\sum_{i=1}^lM_i\right)\max_{r\in[l],s\in[M_r]}\left|(\mu_{rs}(t))_{\eta-\lambda_{1r}}^{\eta}- (\pi_{rs}(t))_{\eta-\lambda_{1r}}^{\eta}\right|\label{u2}\\
&\le&\sum_{i=1}^lM_i+\Delta_2\left(\sum_{i=1}^lM_i\right)\bar{P}^{\max_{r\in[l]}(\lambda_{1r}-\lambda_{2r})^+}\max_{r\in[l],s\in[M_r]}\left(\left|(\mu_{rs}(t))_{\eta-\lambda_{2r}}^{\eta}- (\pi_{rs}(t))_{\eta-\lambda_{2r}}^{\eta}\right|+1\right)\nonumber\\\label{u3}\\
&=&c_1+c(\bar{P})\max_{r\in[l],s\in[M_r]}\left(\left|(\mu_{rs}(t))_{\eta-\lambda_{2r}}^{\eta}- (\pi_{rs}(t))_{\eta-\lambda_{2r}}^{\eta}\right|+1\right)\label{u33}
\end{eqnarray}
where $c_1$ and $c(\bar{P})$ are defined as $\sum_{i=1}^lM_i$ and $c_1\Delta_2\bar{P}^{\max_{r\in[l]}(\lambda_{1r}-\lambda_{2r})^+}$, respectively. \eqref{u1} follows from \eqref{u0} and \eqref{u00}. \eqref{u2} is true as the magnitudes of all the members of $\mathcal{G}$ are less than $\Delta_2$. \eqref{u3} is concluded as for any $X\in\mathcal{X}_{\eta}$ and any $0\le\lambda\le\eta$, we have $\lfloor \frac{X}{\bar{P}^{\lambda}}\rfloor=(X)_{\lambda}^{\eta}$, see Definition \ref{powerlevel}.

So, from \eqref{u000} and \eqref{u33} the probability of alignment of $\bar{\nu}^{[T]}$ with $\breve{\nu}^{[T]}$ is bounded in terms of  $\bar{\nu}(t)$ and $\breve{\nu}(t)$ as follows
\begin{eqnarray}
P_a(\bar{\nu}^{[T]})&\le&\prod_{t=1,c_1+c(\bar{P})<\left|\bar{\nu}(t)-\breve{\nu}(t)\right|}^T\frac{2c_1c(\bar{P})f_{\max}}{\left|\bar{\nu}(t)-\breve{\nu}(t)\right|-c_1-c(\bar{P})}\label{u4}.
\end{eqnarray}

\subsubsection{Bounding the Average Size of Aligned Image Sets}
From (\ref{jen}), we have to compute the following summation,
\begin{eqnarray}
\mbox{E}_{\mathcal{G}}\{|S_{\nu^{[T]}}(\bar{W},\mathcal{G})|\mid \bar{W}=w\}=\sum_{\bar{\nu}^{[T]}\in\mbox{supp}(\breve{U}_{11}^{[T]})}P_a(\bar{\nu}^{[T]})\label{u4,}
\end{eqnarray}
Starting from \eqref{u4,} we have,
\begin{eqnarray}
&&\mbox{E}_{\mathcal{G}}\{|S_{\nu^{[T]}}(\bar{W},\mathcal{G})|\mid \bar{W}=w\}\nonumber\\
&\le&\sum_{\bar{\nu}^{[T]}\in\mbox{supp}(\breve{U}_{11}^{[T]})}\prod_{t=1,c_1+c(\bar{P})<\left|\bar{\nu}(t)-\breve{\nu}(t)\right|}^T\frac{2c_1c(\bar{P})f_{\max}}{\left|\bar{\nu}(t)-\breve{\nu}(t)\right|-c_1-c(\bar{P})}\label{u5}\\
&=&\prod_{t=1}^T\left(\sum_{|\bar{\nu}(t)|\in \{0,1,2,\cdots,c_1\Delta_2\bar{P}^{\max_{j\in[l]}\lambda_{1j}}\},\left|\bar{\nu}(t)-\breve{\nu}(t)\right|\le c_1+c(\bar{P})}1\right.\nonumber\\
&&\left.+\sum_{|\bar{\nu}(t)|\in \{0,1,2,\cdots,c_1\Delta_2\bar{P}^{\max_{j\in[l]}\lambda_{1j}}\},c_1+c(\bar{P})<\left|\bar{\nu}(t)-\breve{\nu}(t)\right|}\frac{2c_1c(\bar{P})f_{\max}}{\left|\bar{\nu}(t)-\breve{\nu}(t)\right|-c_1-c(\bar{P})}\right)\label{u6}\\
&\le&\prod_{t=1}^T\left(2c_1+2c(\bar{P})+1+4c_1c(\bar{P})f_{\max}\left(1+\ln\left(c_1\Delta_2\bar{P}^{\max_{j\in[l]}\lambda_{1j}}\right)\right)\right)\label{u7}\\
&=&{\left(2c_1+2c(\bar{P})+1+4c_1c(\bar{P})f_{\max}\left(1+\ln\left(c_1\Delta_2\bar{P}^{\max_{j\in[l]}\lambda_{1j}}\right)\right)\right)}^T\label{u8}
\end{eqnarray}
Note that from $\eqref{u0}$, $|\bar{\nu}(t)|\in \{0,1,2,\cdots,c_1\Delta_2\bar{P}^{\max_{j\in[l]}\lambda_{1j}}\}$. \eqref{u6}  follows from interchange of the summation and the product \footnote{ Note that from \cite{Arash_Jafar_sumset} for the arbitrary functions $f_1(x),f_2(x),\cdots,f_T(x)$ and the arbitrary sets of numbers $S_1,S_2,\cdots,S_T$ we have,
\begin{align}
&\sum_{a_1\in S_1,a_2\in S_2,\cdots,a_T\in S_T}\prod_{t=1}^Tf_t(a_t)\nonumber\\
=&\sum_{a_1\in S_1}\sum_{a_2\in S_2}\cdots\sum_{a_T\in S_T}\prod_{t=1}^Tf_t(a_t)\\
=&\sum_{a_1\in S_1}f_1(a_1)\times\sum_{a_2\in S_2}f_2(a_2)\times\cdots\times\sum_{a_T\in S_T}f_T(a_T)\label{ret}\\
=&\prod_{t=1}^T\sum_{a_t\in S_t}f_t(a_t)
\end{align}}. \eqref{u7} is true as  for any positive integer number $m$, $\sum_{i=1}^m\frac{1}{i}\le1+\ln(m)$.
\subsubsection{Combining the Bounds to Complete the Proof}
From \eqref{jen} and \eqref{u8}, we have,
\begin{eqnarray}
&&H(\breve{U}_{11}^{[T]}\mid W,W_1,\mathcal{G})-H(\breve{U}_{21}^{[T]}\mid W,W_1,\mathcal{G})\nonumber\\
&\le&\max_{w\in\mathcal{W}}\log{\left\{\mbox{E}_{\mathcal{G}}\{|S_{\nu^{[T]}}(\bar{W},\mathcal{G})|\mid \bar{W}=w\}\right\}}\\
&\le&T\log{\left(2c_1+2c(\bar{P})+1+4c_1c(\bar{P})f_{\max}\left(1+\ln\left(c_1\Delta_2\bar{P}^{\max_{j\in[l]}\lambda_{1j}}\right)\right)\right)}\\
&\le&T~{\max_{r\in[l]}(\lambda_{1r}-\lambda_{2r})^+}\log{\bar{P}}+T~o(\log{\bar{P}})
\end{eqnarray}
Recall that, $c(\bar{P})$ was defined as  $c_1\Delta_2\bar{P}^{\max_{r\in[l]}(\lambda_{1r}-\lambda_{2r})^+}$.

\bibliographystyle{IEEEtran}
\bibliography{Thesis}
\end{document}